\documentstyle{mn}
\newcommand{\mincir}{\ \raise -2.truept\hbox{\rlap{\hbox{$\sim$}}\raise5.truept
	\hbox{$<$}\ }}			
\newcommand{\magcir}{\ \raise -2.truept\hbox{\rlap{\hbox{$\sim$}}\raise5.truept
 	\hbox{$>$}\ }}		
\def\lya{Ly$\alpha$~}
\def\etal{{\it et~al.\ }}
\def\refer{\par\noindent\hangindent 20pt}
\begin{document}

\title[PROPERTIES OF THE LYMAN ALPHA CLOUDS]{PROPERTIES OF THE LYMAN
ALPHA CLOUDS FROM NON-EQUILIBRIUM PHOTOIONIZATION MODELS}
\author[Ferrara \& Giallongo]{
A. Ferrara$^{1}$ {\rm and} E. Giallongo$^{2}$\\
${^1}$ Osservatorio Astrofisico di Arcetri,
Largo E. Fermi 5, I-50125 Florence, Italy\\
${^2}$ Osservatorio Astronomico di Roma, 
via dell'Osservatorio, I-00040 Monteporzio, Italy
}
\maketitle

\begin{abstract}
We investigate the thermal and ionization history of \lya clouds
photoionized by a time--dependent UV background, including
nonequilibrium effects.  The results show that it is possible to
obtain temperatures as low as $T\sim 15000$ K (or, equivalently,
Doppler parameters $b\simeq 15$~km~s$^{-1}$) at $z=3$ for cloud total
densities $n\sim 10^{-4}$ cm$^{-3}$, if (i) the reionization epoch
occurred at $z_i \sim 10$, and (ii) the UV background has a factor
$70-100$ decrease at the HeII edge. A trend towards smaller $b$
with increasing redshift is present in the redshift interval $z=1-5$.
Higher densities lead to higher
values of $b$ and smaller hydrogen correction factors,
$n_{HII}/n_{HI}$. The correction factors for helium are also given.
For a hydrogen column density $N_{HI}=3\times 10^{14}$~cm$^{-2}$,
cloud sizes are larger than 100 kpc, consistent with recent
observations of quasar pairs. Pressure confined models, instead, yield
implausibly low cloud densities at low redshift, and too small sizes
at intermediate redshift. The implications of the model are confronted
with the available observational data.
\end{abstract}
\nokeywords
\section{Introduction}

The study of the \lya absorbers seen in quasar spectra shortward of
the quasar Ly$\alpha$ emission has a considerable cosmological 
relevance as a probe of the physical properties of the neutral gas phase
of the baryonic component of the Universe.

The optically thin Ly$\alpha$ lines are usually associated to highly ionized
clouds ($n_H/n_{HI}\sim 10^4$) in photoionization equilibrium with the 
general UV ionizing background (UVB) produced by quasars. 
The clouds may reside in the intergalactic space between galaxies or
be associated with the big halos surrounding high redshift galaxies
(Bahcall \& Spitzer 1969; Sargent \etal 1980; Lanzetta \etal 1994).

The knowledge of the physical and cosmological properties of the Ly$\alpha$
absorbers relies on a few parameters (redshift, column density, Doppler
width) which can be derived from the line-profile fitting of  (often blended)
absorption features.

The column density distribution is a double power-law with a break at 
$\log N_{HI}=14$ below which the slope is $\beta _f\sim 1.4$.
Above the break the slope is steeper, $\beta _s\sim 1.8$
(Giallongo \etal 1996).

High resolution data taken at resolution $R\sim 30000$ seem to
indicate typical Doppler values of the order of $b\sim 23-25$ km
s$^{-1}$ corresponding to cloud temperatures $T\sim 32000-38000$ K
(Cristiani \etal 1995, Hu et al. 1995, Giallongo \etal 1996).

It is important to note that narrow absorption lines with $b\mincir
20$ km s$^{-1}$ have been observed in high resolution Ly$\alpha$
samples.  Chaffee \etal (1983) first noted that a line in their
spectrum at resolution $\sim 12$ km s$^{-1}$ was surprisingly narrow
with $b\simeq 15$ km s$^{-1}$. More recent echelle data with variable
signal-to-noise ratios reveal a non negligible fraction of narrow
lines with Doppler parameter in the range $10<b<20$ km s$^{-1}$. The
few lines ($\sim 1$\%) with $b\mincir 10$ km s$^{-1}$ present in the
spectra have been often identified as metal-lines associated with
heavy element systems (e.g. Rauch \etal 1993, Hu et al. 1995).

It is difficult to derive the intrinsic fraction of narrow lines with
$b\sim 15-20$ km s$^{-1}$ since it depends on the S/N ratio,
intrinsic blending of the lines, and resolution
of the various spectra and on the fitting procedure adopted. These fractions
range from 10\% to 20\% of the lines with $\log N_{HI}\magcir 12.8$
(Carswell \etal 1991; Rauch \etal 1993; Cristiani \etal 1995; Hu et al. 1995;
Tytler 1995; Giallongo \etal 1996).

In any case, reconciling temperatures as low as $T\simeq 15000$ K with
sizes along the l.o.s. greater than 30-50 kpc for the Ly$\alpha$
clouds proves to be a difficult challenge.  If the clouds are assumed
in photoionization and thermal equilibrium with a power-law UV
background produced by quasars at $z=2-4$, then temperatures $T <
30000$ K could only be obtained increasing the hydrogen volume density
to $n_H > 10^{-3}$ cm$^{-3}$. As a consequence, cloud sizes along the
line of sight drop down to $S_{\|}<100$ pc, as outlined by Chaffee
\etal (1983), Pettini \etal (1990), and Donahue \& Shull (1991).

Besides, a growing number of  Ly$\alpha$ forest observations in quasar pairs
suggest large transverse sizes of the clouds. Foltz \etal (1984)
derived a lower limit in the range $5-25 ~h_{50}^{-1}$ kpc for the
gravitational lensed pair Q2345+0007, while Smette \etal (1992) gave 
lower limits in the range $12-50 ~h_{50}^{-1}$ kpc for the QSO pair UM673. 
Recent measures by Bechtold \etal (1994)
give larger values with a 90\% lower limit $80 h_{50}^{-1}$ kpc 
at $z=1.8$ for the diameters of
spherical clouds.  Smette etal (1995) find a $2\sigma$ lower
limit of $100 h_{50}^{-1}$ kpc for the diameter of spherical
clouds at $z\sim 2$.

Given the implied low densities, radiative recombination can not be 
the main cooling agent for clouds with $T\mincir 20000$ K. 
Duncan, Vishniac \& Ostriker (1991)
proposed adiabatic cooling due to cloud expansion against the external
pressure of a diffuse but hot ($T>10^5$ K) intergalactic medium. They
obtained cloud temperatures $T\sim 15000$ K with sizes $\sim 20$ kpc.
However, their calculations are based on the assumption of ionization 
equilibrium, and neglect heating from helium photoionization, which 
increases the cloud temperature by a factor 1.8.

Giallongo \& Petitjean (1994) (GP) stressed the importance of the
inverse Compton cooling on the CMBR, and of a steepening of the
ionizing UV background at wavelengths shorter than the HeII edge. The
combination of the two effects was shown to produce temperatures
$T\sim 22000$ K with sizes up to 100 -- 200 kpc.

However, GP imposed both thermal and ionization equilibrium for clouds
with densities $n_H\sim 10^{-4}$ cm$^{-3}$ at $z\sim 3$, an assumption
which is only marginally consistent with the relevant timescales of
the problem.

Models which include the effect of gravity have also been proposed
(Petitjean et al. 1994; Meiksin 1994) in the context of the so
called minihalo model (Rees 1986). Typical Doppler values $>20$ km s$^{-1}$
are found together with cloud sizes $<60$ kpc.

In the last period, several authors have studied the properties of the
Ly$\alpha$ clouds in the general cosmological cold dark matter scenario
(Hernquist et al. 1995; Zhang et al. 1995; Miralda-Escud\'e et al. 1995).

Although these models follow the formation and evolution of the
Ly$\alpha$ clouds in a dynamical scenario, the treatment of
photoionization is only approximate either because of the assumption
of ionization equilibrium or because of the uniform ionizing
background adopted with fixed initial conditions.

It is important to investigate time dependent photoionization models 
for the Ly$\alpha$ clouds for the following reasons: (i) the UVB produced
by quasars has a strong dependence on redshift; (ii) the 
inverse Compton cooling time, $t_c$, is longer than the Hubble time, $t_H$, at
the typical redshifts at which Ly$\alpha$ clouds are observed:  $t_c/t_H\sim
40 (1+z)^{-2.5}$ (GP).          

The aim of this study is to investigate in detail the thermal and
ionization history of the clouds through a nonequilibrium
photoionization model. This simple model allows us to explore the
ionization state of the cloud as a function of a time--dependent UV
background with different reionization epochs and with different
shapes in frequency near the HeII edge.

Our results show that it is possible to obtain temperatures as low as 
$T\sim 15000$ K, for cloud total densities $n\sim 10^{-4}$ cm$^{-3}$ 
if the reionization epoch occurred at $z_i \magcir 10$ and
if the UV background has a factor $70-100$ decrease at the HeII edge.

\section{Photoionization Model}
The equations governing the thermal and ionization evolution  of a \lya cloud
immersed in an isotropic ionizing background of intensity $I_{\nu}$ are (see Ferrara
\& Field 1994 for details):

$${d\over dt}n_{X^{i+1}} =n_{X^i}\gamma_p(X^i) + \gamma_c(X^i,T)
n_{X^i} n_e - n_{X^{i+1}} n_e \alpha (X^i,T),\eqno(2.1)$$
 
$${k\over (\gamma-1)} {d\over dt}\left[ T \sum_X {h_X\over \mu_{X,i}}\right] = - 
{1\over n}{\cal L}(X^i)+ {p\over n^2}  {dn\over dt}.\eqno(2.2)$$

The gas is assumed of primordial composition with a helium abundance
equal to $n_{He}=0.1 n_H$ and total number density $n$; even if metals 
are present, they do not affect the temperature very much provided the
ionization level is high, as the one found here.

We consider the evolution of
the following species: $H^0, H^+, He^0, He^+, He^{++},$ and electrons. 
The symbol $X= H, He$ denotes the element considered,  
and $i$ its state of ionization; obviously,
$i=0$ for hydrogen, and $i=0,1$ for helium; $n_e$ is the electron density,
$\gamma$ is the specific heat ratio assumed to be constant and equal to 5/3.

The photoionization rate $\gamma_p$ is given by 
$$\gamma_p(X^i)=\int_{\nu_{L_X}}^{\infty} {J_{\nu}\over h \nu}
\sigma_\nu(X^i) d\nu \left[1+\phi(X^i)\right]$$
$J_\nu$ is the first moment of the field, $\nu_{L_X}$ indicates the
ionization limit for each species,
$\sigma$ is the photoionization cross-section,  $\phi$ is the
secondary ionization rate, $\gamma_c$ is the collisional ionization
coefficient, and $\alpha$ is the total recombination coefficient.
We have adopted the ``on the spot'' approximation in which
the diffuse field photons are supposed to be absorbed close to the point
where they have been generated.

The energy equation (2.2) governs the evolution of the gas temperature $T$.
We have used the following notation:  $h_X=n_X/n$ is the relative abundance 
of the element $X$; $x_{X,i+1}=n_X^{i+1}/n_X$ is the fractional density of the
ionization state $i+1$ of the element $X$; 
$\mu_{X}=(i_{max}+1)^2[1+\sum_i^{i_{max}}  (i+1) x_{X,i+1}]^{-1}$,
with $i_{max}=0,1$ for H and He, respectively.

The function ${\cal L}$ represents the net cooling rate per unit volume and is
given by the difference between photoionization heating and radiative cooling:

$$ {\cal L}(X^i)=\sum_i \sum_X n(X^i)\int_{\nu_{L_X}}^{\infty} {J_{\nu}\over h \nu}
h[\nu-\nu_{L_X}]\sigma_\nu(X^i) d\nu - \sum_i \sum_X  \Lambda
(X^i).$$

The following processes have been included in the calculation of the
cooling function: (i) free-free from all ions; (ii) H and He recombinations;
(iii) electron impact ionization of H and He; (iv) electron impact excitation
of H and He (n=2,3,4 triplets); (v) He dielectronic recombination. At high 
temperatures or early epochs inverse Compton cooling on the CMWB photons is 
important and it has been included as in Ikeuchi \& Ostriker (1986).

We mostly concentrate on constant cloud total density cases,
(i.e., $dn/dt = 0$ in eq. [2.2]); in addition, we also explore cases in which 
\lya clouds are taken to be in pressure equilibrium with (and hence confined by) the 
surrounding IGM. To this aim, it is useful to  write eq. (2.2) as follows:

$${\gamma\over (\gamma-1)}k {d\over dt}\left[ T \sum_X {h_X\over \mu_{X,i}}\right] = 
- {1\over n }{\cal L}(X^i)+ {kT\over p} \sum_X {h_X\over \mu_{X,i}} {dp_{I}\over dt}, 
\eqno(2.3)$$

where $p_{I}(z)$ is the pressure of the IGM, which is a prescribed function
of the redshift $z$. Ikeuchi \& Ostriker (1986)  (and more recently
Shapiro 1995) have studied in detail the thermal history of the IGM
from the era of the reheating to the present; they conclude that the most
plausible scenario requires that both photoionization and bulk mechanical
heating (i.e. shocks) contribute to the heating. Ikeuchi \& Ostriker
show that, if the reionization occurred at $z_i=10$ - for example - the 
initial evolution of the IGM is practically isothermal,  $p_I \propto (1+z)^3$,
and the IGM enters the adiabatic expansion phase, $p_I \propto (1+z)^5$, 
at $z\mincir 4$. As for the local value of the pressure, we take  $p_I(0)= 
10^{-2}$~cm$^{-3}$~K; we will use the cosmology $\Omega=1$ and $h_{50}=1$.  

The problem is completely specified once the evolution of the UVB field is 
assigned together with the initial conditions (the latter are discussed below).
We have assumed an evolving UVB following Madau (1991) and
Meiksin \& Madau (1993) model. The intensity of the UVB at the Lyman limit,
$J_{LL}$, evolves as $(1+z)^3$ for $z< 2$; it is approximately
constant, i.e. $\propto (1+z)^{1/4}$, in the range $2\leq z \leq 3$, and it
declines as $\propto \exp[-0.69(z-3)]$ for $z>3$.

We assume a normalization value $J_{LL}=7\times 10^{-22}$ erg s$^{-1}$ 
cm$^{-2}$ sr$^{-1}$ Hz$^{-1}$ at $z=2$, consistent with the
average value $J_{LL}=5\times 10^{-22}$ found from the proximity effect
of a Ly$\alpha$ high resolution sample evaluated in the redshift range 
$z=1.7-4.1$ (Giallongo \etal 1996).
A power-law spectrum with index $\alpha =-1.5$ for the UVB has been adopted.

In some cases a factor 100 break has been introduced at 4 Ryd to simulate the 
HeII absorption trough, consistently with the detections of the HeII Ly$\alpha$ 
Gunn-Peterson effect which have provided an indirect evidence of a steepening 
at the HeII edge with a UVB ratio between the Lyman limit and the HeII edge 
$S_L \sim 100$ (Jakobsen \etal 1994).

\section{Model Results}

Using the photoionization model described in the previous section we
have first explored the thermal and ionization history of clouds having 
constant total density. This isochoric scenario, although unrealistic
if followed for the overall Hubble time, allows a simple insight on 
the evolution of the ionization and thermal state of low density
clouds in response to changes of the UVB.

Among the several cases investigated, we present the most relevant ones 
featuring (i) different initial conditions

\begin{figure}
\vspace{150mm}
\caption{ 
(a) Evolution of the Doppler perameter $b$ (lower panel) and
of the hydrogen correction factor $\chi_{H,1}$ (upper panel) for
constant cloud density $n=10^{-4}$~cm$^{-3}$ as a function of redshift
for different initial conditions (see text). The initial conditions
are $z_i=7,10$ and $T_i=10^4, 10^5$~K. Points refer to $b$ obtained
assuming ionization and thermal equilibrium for comparison assuming
the same UVB.  {\it Solid} lines indicate cases with $B=1$ (no break);
{\it dotted} lines indicate cases with $B=100$. The corresponding
temperature scale in the lower panel is shown on the right axis.
(b) As Fig. 1a, with initial conditions $z_i=7$ and $T_i=10^4$~K,
$B=1$ for different values of the density: $n=10^{-3}$~cm$^{-3}$
(dotted line), $n=10^{-4}$~cm$^{-3}$ (solid), $n=10^{-4.5}$~cm$^{-3}$
(long-dashed).  
}
\label{figure 1a}
\end{figure}

(reionization 
epoch, $z_i=7, 10$, and temperature, $\log T_i=4,5$) and (ii) different 
values of the HeII break at 4 Ryd, $B=1,100$, assuming a total density 
$n=10^{-4}$ cm$^{-3}$. The gas is supposed to be mostly neutral at $z=z_i$; 
different assumptions produce slight changes only for redshifts outside the
currently observable range. At $z=10$ the Compton cooling time is so short 
that there are no relevant    differences between the two different $T_i$.

Fig. 1a shows the thermal and ionization evolution of 
a \lya cloud as described by the Doppler parameter, $b=(2kT/m_H)^{1/2}$
and the hydrogen correction factor, $\chi_{H,1}=n_{HII}/n_{HI}=x_{H,1}
/(1-x_{H,1})$.
The Doppler parameter evolves strongly as a function of redshift and,
independently of $T_i$ and $z_i$, reaches a maximum value $b \simeq 
28$~km~s$^{-1}$ at about $z=1.2$ for $B=1$, i.e., no HeII break; 
the memory of the initial conditions is completely washed away already
at $z\simeq 2$, for both cases $B=1$ and $B=100$. The temperature
peak is due to the lower efficiency of the Compton cooling which decreases
as $(1+z)^4$; for $z \mincir 1.2$ this effect is overwhelmed by the decrease 
of the UVB, and hence of the photoionization heating. The peak position
also lags behind an analogous maximum of $\chi_{H,1}$ (see below) since the 
decrease in the photoionization heating is initially compensated by a
larger neutral fraction.

The cases $B=100$ look qualitatively similar to the $B=1$ ones, but they
show a general tendency towards lower temperatures,
with a minimum value $b\simeq 13$~km~s$^{-1}$ attained at $z=6-7$, 
with $z_i=10$. It is worth mentioning, though, that even at $z=3$ 
typical $b$ values are $16-18$~km~s$^{-1}$, to be compared with $b\simeq
23$~km~s$^{-1}$ for the no break case. 

The hydrogen correction factor $\chi_{H,1}$ traces to a good extent the evolution 
of the UVB, peaking at $z=2$; the range of variation of  $\chi_{H,1}$ is $3000-10^5$.
The evolution is rather insensitive to $z_i$ and $T_i$, the different curves
converging as early as $z\simeq 5$. A $B=100$ break reduces $\chi_{H,1}$ by
$\sim 35\%$ around the maximum ($z=2$).

We have also explored the cases $n=10^{-3}$~cm$^{-3}$ and $n=10^{-4.5}$~cm$^{-3}$
with initial conditions $z_i=7$ and $B=1$; the corresponding evolutionary
curves are shown in Fig. 1b.
The behavior $\chi_{H,1}$ is qualitatively the same as above; in particular
the value of the maximum increases almost linearly with $n$. The Doppler
parameters are lower that for the standard $n=10^{-4}$~cm$^{-3}$ case both
for $n=10^{-3}$~cm$^{-3}$ and for $n=10^{-4.5}$~cm$^{-3}$ (this is due to a 
transition from Compton cooling dominated losses at low $n$ to recombination
dominated cooling at higher $n$). In addition, the maximum of the $b$ curve
shifts towards lower redshifts as the density is decreased.

In general, we can define correction factors for helium in analogy to
the hydrogen one as $\chi_{He,i+1}=n_{He,i+1}/n_{HI}=h_{He}
x_{He,i+1}/h_H(1-x_{H,1})$ (notice the definition relative to neutral
hydrogen). Fig. 2 shows the behavior of the helium correction factors
as a function of redshift. The most relevant feature is the change in
the evolution of the HeII fraction from the cases $B=1$ to the cases
$B=100$. For $B=1$ and $z<5$, $\chi_{He,1}$ has an almost constant low
value $\sim 20$, i.e., most of the He is in the doubly ionized
state. As $B$ is increased to 100, $\chi_{He,1}$ develops a maximum
around $z=3$ of about 4000, a value similar to that of $\chi_{He,2}$
in the same redshift interval. Thus the ratio of the two correction
factors provides a very sensitive measure (and test) of the 4~Ryd
break.

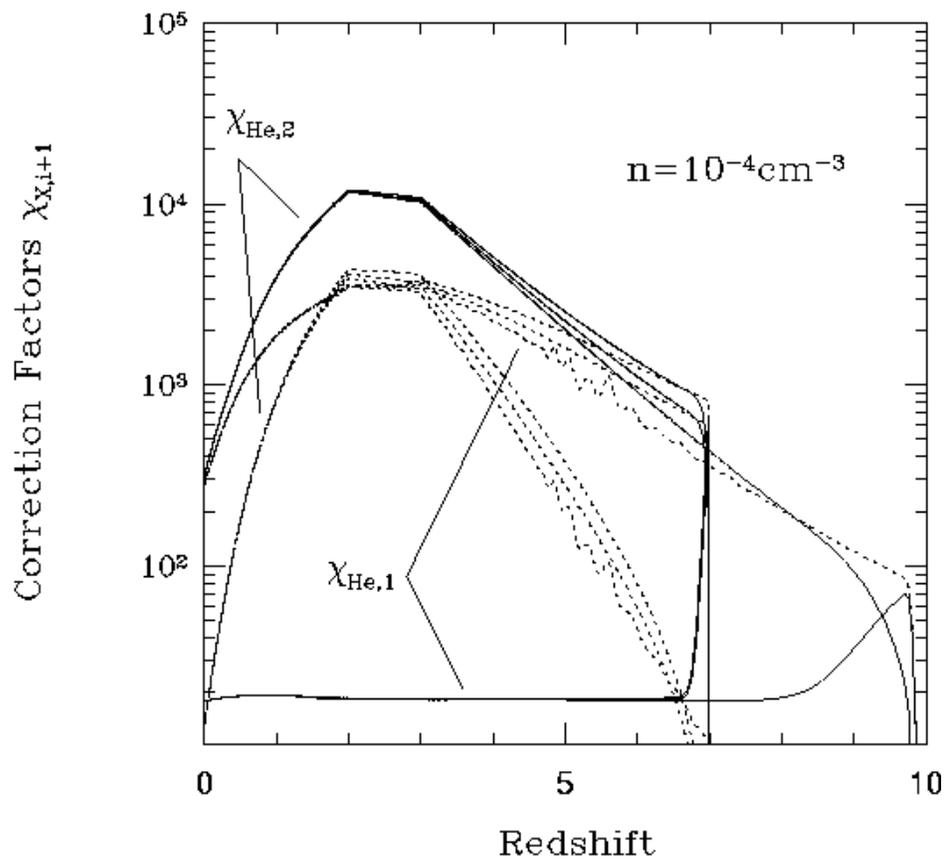
\begin{figure}
\vspace{100mm}
\caption{ Evolution of the helium correction factors, $\chi_{He,1}$
and $\chi_{He,2}$ for constant cloud density $n=10^{-4}$~cm$^{-3}$ as
a function of redshift for the same cases as in Fig. 1a.  {\it Solid}
lines indicate cases with $B=1$ (no break); {\it dotted} lines
indicate cases with $B=100$.  
}
\label{fig2}
\end{figure}

The previous results allow to compute the \lya cloud sizes along the
l.o.s. $S_{\|}=N_{HI} (1+ \chi_{H,1})/n$ at any given redshift, as
illustrated by Fig. 3.  Typical sizes of the clouds are several kpc,
ranging from a minimum of about 5~kpc for $N_{HI}=10^{13}$~cm$^{-2}$
to 100-120 kpc for $N_{HI}=3 \times 10^{14}$~cm$^{-2}$ in the redshift
interval 2-3. Given the higher neutral fraction found in models with
$B=100$, sizes are a factor $\sim 1.5$ smaller.  Apart from $B$, the
sizes are quite insensitive to the initial conditions.  Locally
($z=0$), clouds with $N_{HI}<3\times 10^{14}$~cm$^{-2}$ are predicted
to have sizes not exceeding 5 kpc.

Finally, models in which \lya clouds are assumed to be in pressure
equilibrium with the IGM, according to eq. (2.3), are presented in
Fig. 4. The temperature of the clouds is now a monotonically
increasing function of time, as the cooling reduction caused by the
density decrease dominates the opposite effect due to the decrease of
the photoionization energy input. At $z=3$, $b=23-25$~km~s$^{-1}$,
depending on $B$. We point out that these results depend on the
particular assumption made for the thermal history of the IGM,
following Ostriker \& Ikeuchi (1986); deviations from these values may
produce quantitative changes in the determination of $b$. The same
behavior is shown by $\chi_{H,1}$, which reaches very high values
($\sim 3\times 10^6$) at $z=0$. The total density $n$ of the cloud
drops very rapidly from the initial value $n_i \simeq
10^{-2}$~cm$^{-3}$ to an extremely low value $\simeq
10^{-7}$~cm$^{-3}$ at $z=0$ (essentially the IGM density). As a
consequence, sizes are implausibly large, or, stated in a different
way, the clouds would dissolve rapidly below $z \sim 2$. At
intermediate redshifts, pressure confined models yield essentially the
same $b$ values as the constant density, $B=1$ cases (but higher than
$B=100$ ones); 

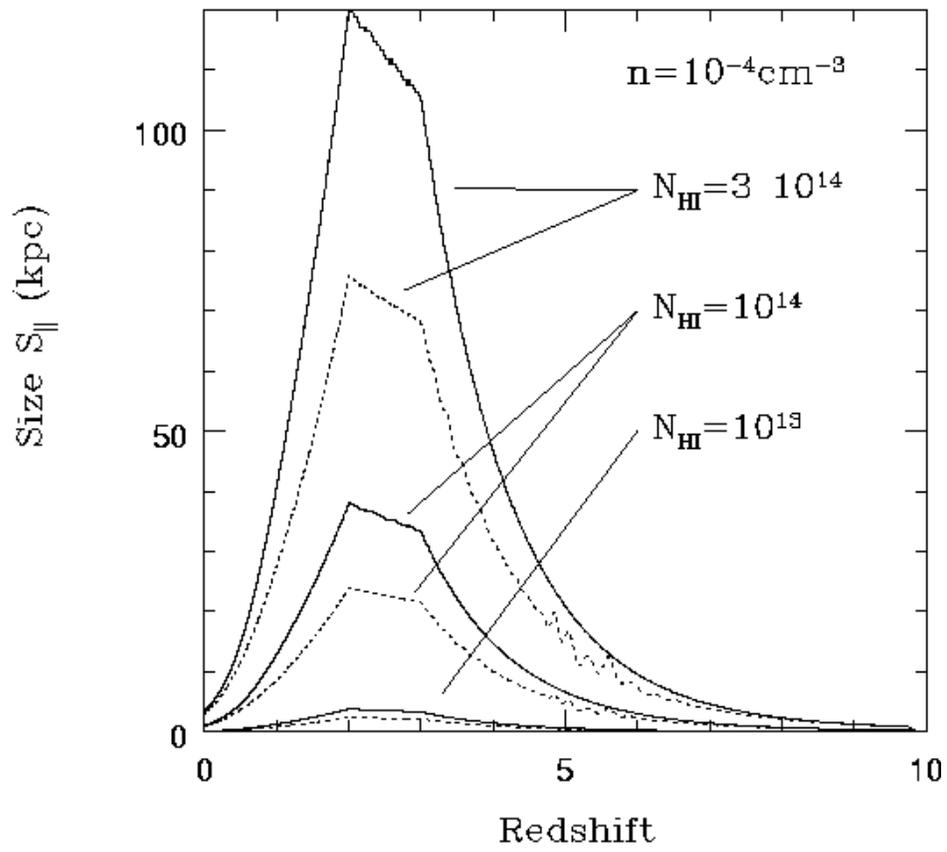
\begin{figure}
\vspace{100mm}
\caption{ 
Cloud sizes $S_{\|}$ for constant cloud density 
$n=10^{-4}$~cm$^{-3}$ as a function of redshift for initial conditions
$z_i=10$, $T_i=10^4$~K. 
{\it Solid} lines correspond to $B=1$ (no break); {\it dotted} lines
correspond to $B=100$. Three different hydrogen column densities $N_{HI}
=10^{13}, 10^{14}, 3\times 10^{14}$~cm$^{-2}$ are shown for each case.
}
\label{fig3}
\end{figure}

however, correction factors (and thus sizes) are
typically one order of magnitude smaller. Thus, low-$z$ \lya clouds
observations can be used to put strong constraints on pressure
confinement models.

\section{Discussion and Implications}                                  

We first compare our time-dependent results with the ones obtained by
GP who assumed thermal and ionization equilibrium. GP used a value of
$J_{LL}= 10^{-21}$ erg s$^{-1}$ cm$^{-2}$ Hz$^{-1}$, a value
corresponding to $z=2.25$ for the presently adopted $J_\nu(z)$. At
that redshift, the cloud temperature (see Fig. 1a) is $T \simeq
3\times 10^4$~K for $B=1$, and $T \simeq 2\times 10^4$~K for $B=100$,
to be compared with the analogous values obtained by GP (see their
Fig. 1), $T \simeq 4.7\times 10^4$~K for $B=1$, and $T \simeq 3\times
10^4$~K for $B=100$. We have plotted in Fig. 1a the values of $b$
obtained assuming ionization and thermal equilibrium for comparison
assuming the same UVB. The reasons for the discrepancy are: (i)
Compton losses on the CMWB strongly cool the clouds at high redshift;
(ii) the UVB decreases exponentially for $z>3$; and (iii)
nonequilibrium effects. The sizes found by GP are about a factor 1.5
larger for the same $N_{HI}$ and $n$, thus implying a larger hydrogen
correction factor. This is due to the enhanced collisional ionization
produced by their higher temperature.

The bottom line of our study is that it is feasible to reconcile two
apparently conflicting observed properties of the \lya clouds, low
temperatures and large sizes, basically with one assumption, i.e. the
existence of a break in the UVB at the HeII edge. It is easy to show
that \lya clouds become optically thick to radiation above the HeII
Lyman limit for a neutral hydrogen column density $N_{HI}=6.3\times
10^{17}/\chi_{He,1}$~cm$^{-2}$. For the calculated values of
$\chi_{He,1}\sim 20$ ($B=1$), \lya clouds with $N_{HI}> 3\times
10^{16}$~cm$^{-2}$ will substantially absorbe the flux above 4~Ryd,
thus enhancing the value of $\chi_{He,1}$ even further (see Fig.  2
for $B=100$). This well-known (Miralda-Escude \& Ostriker 1990,
Shapiro 1995) nonlinear effect is thus responsible not only for the
break, but also, given the reduced HeII ionizing flux, for a
detectable HeII Gunn-Peterson effect (Jakobsen \etal 1994). In
addition, from our results we do expect the presence of a pronounced
He II \lya forest.

\begin{figure}
\vspace{100mm}
\caption{ 
As Fig. 1a, but for the constant pressure case, and initial 
conditions $z_i=7, 10$,  $T_i=10^4$~K. The {\it dashed} line in the upper panel
is the total density, $n$, of the cloud.
{\it Solid} lines correspond to $B=1$ (no break); {\it dotted} lines
correspond to $B=100$. 
}
\label{fig4}
\end{figure}

The issue about Doppler parameters and cloud sizes deserves some
additional discussion. First it appears that $b$ values lower than
about 15~km~s$^{-1}$ cannot be reproduced by any of our nonequilibrium
models.  This agrees with the presence of a lower envelope of the
$b$-distribution with a cut-off at $b_c\simeq 16$~km~s$^{-1}$ found by
different investigators (Carswell \etal 1991, Rauch \etal 1992, 1993,
Hu \etal 1995, Giallongo \etal 1996), disputing a previous claim by
Pettini \etal (1990) of a lower value of $b_c$, probably affected by
line selection and fitting procedure artifacts. Quite interestingly,
we find also an upper limit $b\simeq 28$~km~s$^{-1}$, whereas the
above studies have inferred $b$-values as high as
$60$~km~s$^{-1}$. The most natural suggestion is that high-$b$ values
can be unresolved blends of $b\simeq b_c$ lines (Rauch \etal 1992, Hu
\etal 1995). However, in this case the two-point correlation function
should display an increased amplitude on small scales, typically
$\Delta v<100$~km~$s^{-1}$. Some indirect support to this hypothesis
is lent by Cristiani \etal 1995 and Meiksin \& Bouchet (1995), who
both found significant clustering on scales $\Delta v \ge
100$~km~$s^{-1}$. If blending is viable, depatures from Maxwellian
velocity distribution are expected (Kulkarni \& Fall 1995). A
different explanation, i.e. bulk motions, has been shown (Press \&
Rybicki 1993) to encounter apparently insurmountable
difficulties. Alternatively, if a hot IGM does exist, temperatures of
the order of a few $\times 10^5$~K could be generated in the
interfaces between the IGM and the clouds by thermal conduction
(Ferrara \& Shchekinov 1996).

The cloud sizes $S_{\|}$ derived here for $n=10^{-4}$~$cm^{-3}$,
albeit larger than in previous models, are still smaller than the
transverse ones, $S_\perp$, inferred either from lensed or paired
quasars. Thus, unless cloud densities are smaller than $\sim
10^{-4}$~$cm^{-3}$, we are forced to conclude that clouds must be
flattened, a conclusion shared by other investigators on different
grounds (Rauch \& Haehnelt 1995, Hu \etal 1995). Dinshaw \etal
(1994,1995) have provided very interesting evidence for the evolution
of $S_\perp$: from their quasar pair experiments they obtain $S_\perp>
320 h^{-1}_{50}$~kpc in $z=0.5-0.9$ and $S_\perp> 80 h^{-1}_{50}$~kpc
at $z=1.8$. From Fig. 3, independently of $N_{HI}$, the ratio
$S_{||}(z=1.8)/S_{||}(z=0.5)= 5.6$, while the analogous ratio for
$S_\perp$ is $\sim 0.25$. Although very speculative, this finding
seems to suggest that the clouds evolve from an approximately round
shape to a highly flattened one, reminiscent of a gravitational
collapse, between $z\sim 2$ and $z\sim 1$.
          
\bigskip
\noindent
{\bf REFERENCES}
\bigskip
      
\refer {Bahcall, J. N. \& Spitzer, L. 1969, ApJ, 166, L63}

\refer {Bechtold, J., Crotts, A. P. S., Duncan, R. C, \& Fang, Y. 1994, ApJ, 437,
L83}

\refer {Carswell, R. F., Lanzetta, K. M., Parnell, H. C., \& Webb,
J. K. 1991, ApJ, 371, 36}

\refer {Chaffee, F.H. Jr., Weymann, R. J., Latham, D. W. \& Strittmatter,
P. A. 1983, ApJ, 267, 12}

\refer {Cristiani, S., D'Odorico, S., Fontana, A., Giallongo, E.
\& Savaglio, S. 1995 , MNRAS, 273, 1016}

\refer  {Dinshaw, N., Impey, C. D., Foltz, C. B., Weymann, R. J. \& Chaffee, F. H.
1994, 437, L87}

\refer  {Dinshaw, N., Foltz, C. B., Impey, C. D., Weymann, R. J. \& Morris, S. L. 
1995, Nature, 373, 223}

\refer {Donahue, M., \& Shull, J. M. 1991, 383, 511}

\refer {Duncan, R. C.,  Vishniac, E. T., \& Ostriker, J. P. 1991, ApJ, 368, L1}

\refer {Ferrara, A. \& Field, G. B. 1994, ApJ, 423, 665}

\refer {Ferrara, A. \& Shchekinov, Y. 1996, ApJL, in press }

\refer {Foltz, C. B., Weymann, R. J., Roser, H. J., \& Chaffee, F. H., Jr. 1984,
ApJ, 281, L1}
 
\refer {Giallongo, E. \& Petitjean, P. 1994, ApJ, 426, L61 (GP)}

\refer {Giallongo, E., Cristiani, S., D'Odorico, S., Fontana, A.,
\& Savaglio, S. 1996, ApJ, in press }

\refer {Hernquist, L., Katz, N., Weinberg, D. H., Miralda-Escud\'e, J.
1996, ApJ, 457, L51}

\refer {Hu, E. M., Kim, T., Cowie, L.L., \& Songaila, A. 1995, AJ, 110, 1526}

\refer {Jakobsen, P., Boksenberg, A., Deharveng, J. M., Greenfield, P.,
Jedrzejewski, R. \& Paresce, F. 1994, Nature, 370, 35}

\refer {Lanzetta, K. M., Bowen, D. V., Tytler, D., \& Webb, J. K.  1994,
ApJ, 442, 538}

\refer {Ikeuchi, S. \& Ostriker, J. P. 1986, ApJ, 301, 522}

\refer {Kulkarni, V. P., \& Fall, S. M. 1995, ApJ, 453, 65}

\refer {Madau, P. 1991, ApJ, 376, L33}

\refer {Meiksin, A. 1994, ApJ, 431, 109}

\refer {Meiksin, A., \& Madau, P. 1993, ApJ, 412, 34}

\refer {Meiksin, A., \& Bouchet, F. R.  1995, ApJ, 448, L85}

\refer {Miralda-Escud\'e, J., Cen, R., Ostriker, J. P., \& Rauch, M. 1995,
submitted to ApJ}

\refer {Pettini, M., Hunstead, R. W., Smith, L. J., \& Mar, D. P. 1990, MNRAS, 
246, 545}

\refer {Press, W. H., \& Rybicki, G. B. 1993, 418, 585}

\refer {Rauch, M., Carswell, R. F., Chaffee, F. H., Foltz, C. B., Webb,
J. K., Weymann, R. J., Bechtold, J., \& Green, R. F.  1992, ApJ, 390,
387}

\refer {Rauch, M., Carswell, R. F., Webb, J. K., \& Weymann, R. J. 1993, MNRAS,
260, 589}

\refer {Rauch, M., \& Haehnelt, M. G. 1995, MNRAS, 275, L76}

\refer {Rees, M. J. 1986 MNRAS, 218, 25P}

\refer {Sargent, W. L. W., Young, P. J., Boksenberg, A. \& Tytler, D. 1980, ApJS,
42, 41}

\refer {Shapiro, P. R. 1995, in The Physics of the Interstellar and Intergalactic
Medium, PASP Series, Vol. 80,  eds. A. Ferrara, C. F. McKee, C. Heiles, P. R.
Shapiro, (PASP: San Francisco), 55}

\refer {Smette, A., \etal   1992, ApJ, 389, 39}
 
\refer {Smette, A.,  Robertson, J. G., Shaver, P. A., Reimers,
D., Wisotzki, L. \& K\"ohler, Th. 1995, A\&A Suppl., 113, 199}

\refer {Tytler, D. et al. 1995, in QSO Absorption Lines, Springer, Ed. G. Meylan,
289.}

\refer {Zhang, Yu, Anninos, P., Norman, M. L. 1995, ApJ, 453, L57}
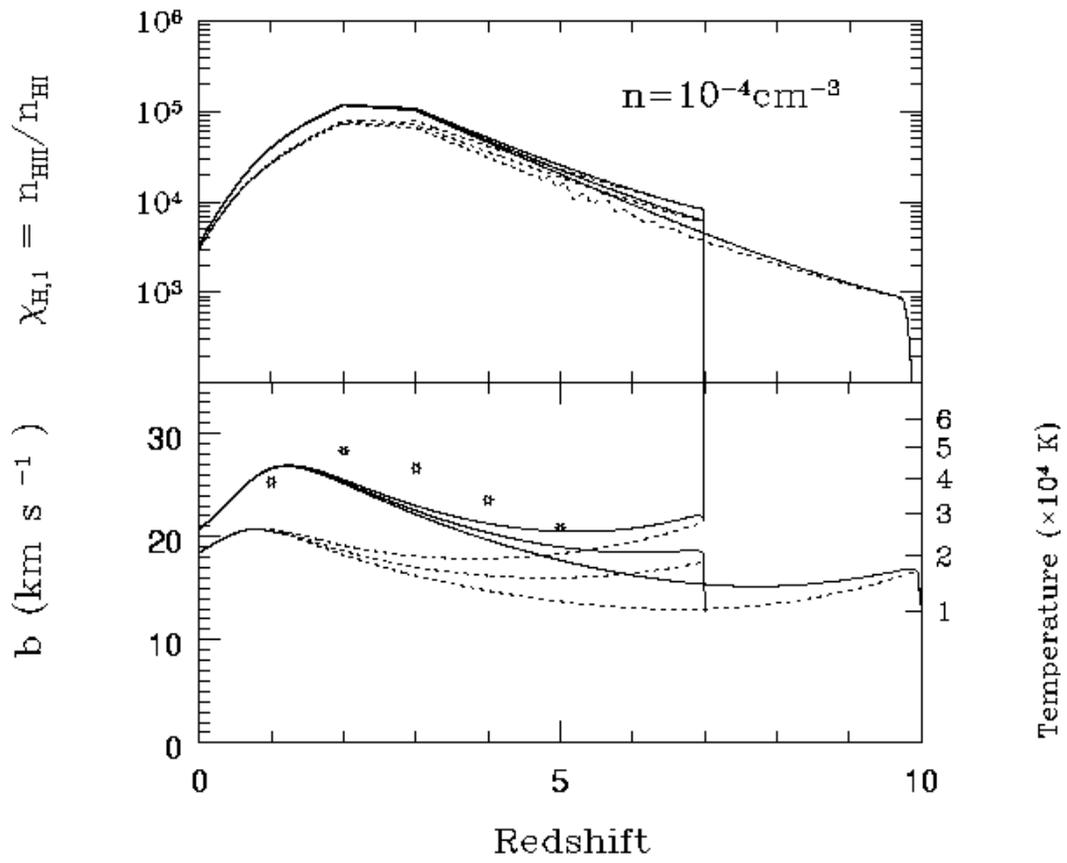
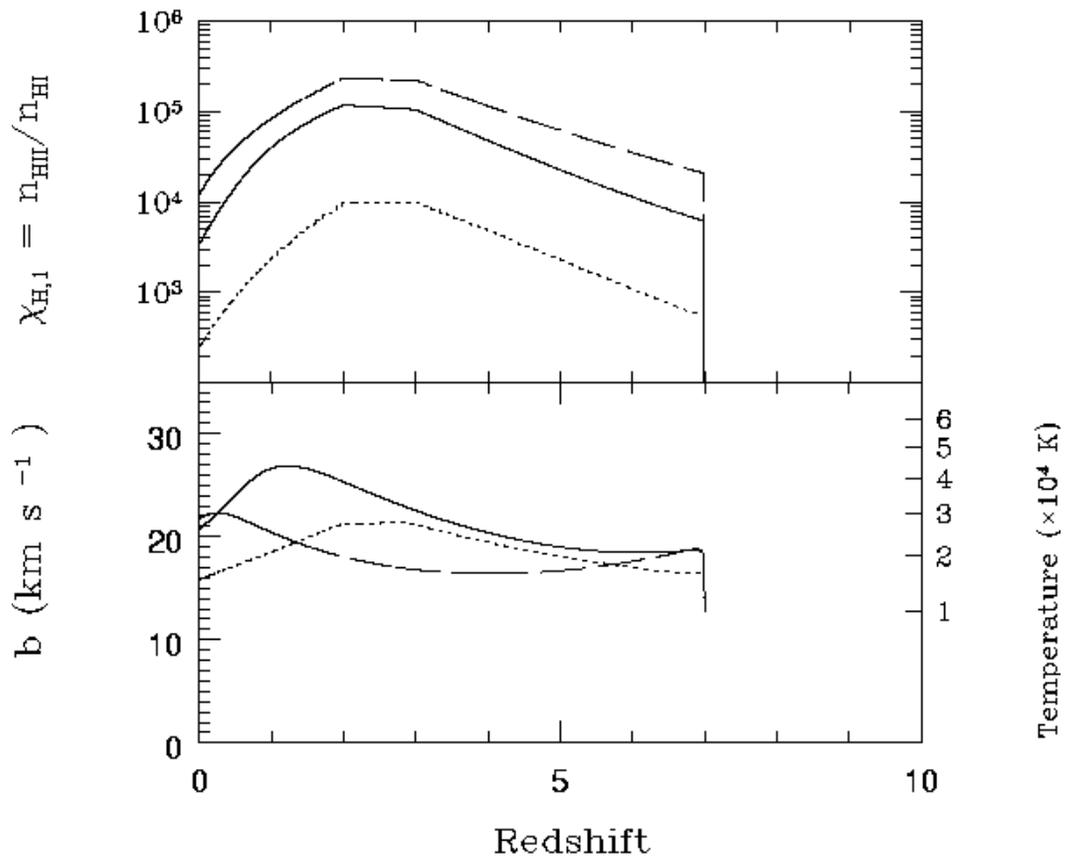
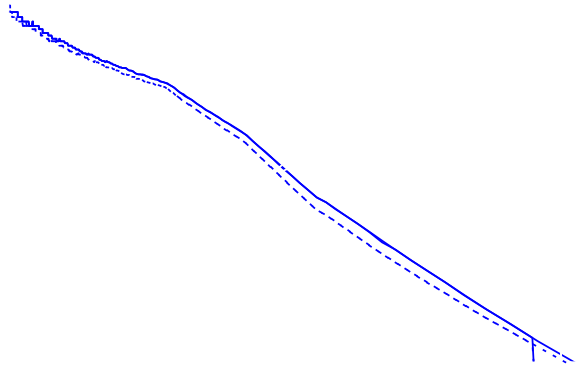
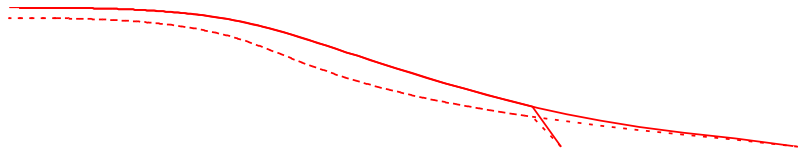
\end{document}